\documentclass[amstex,aps,amsfonts,prsty]{article}

\usepackage{amssymb}
\usepackage{amsmath}
\usepackage{graphicx}
\usepackage{citesort}

\def \be{\begin{equation}}
\def \ee{\end{equation}}
\def \ba{\begin{array}{l}}
\def \Ba{\begin{array}{ll}}
\def \ea{\end{array}}
\def \bq{\begin{eqnarray}}
\def \eq{\end{eqnarray}}
\def \nn{\nonumber\\}
\def \lb{\label}
\def \ln{{\rm ln}}

\def \fr{\frac}

\def \b{\beta}
\def \d{\delta}
\def \D{\Delta}

\def \f{\phi}

\def \G{\Gamma}
\def \lm{\lambda}

\def \n{\nabla}

\def \t{\tau}

\def \tl{\tilde}
\def \ol{\overline}

\def \la{\langle}
\def \ra{\rangle}
\def \[{\left[}
\def \]{\right]}
\def \({\left(}
\def \){\right)}

\def \R{R_{c}}
\def \T{T_{c}}

\def \I{\int d^{D}x}
\def \2{\frac{1}{2}}
\def \4{\frac{1}{4}}

\begin{document}

\begin{center}

{\Large \bf Non-perturbative phenomena in the three-dimensional
            random field Ising model}

\vskip .2in

Vik.S.\ Dotsenko

\vskip .1in

LPTMC, Universit\'e Paris VI,  4 place Jussieu, 75252 Paris, France  

L.D.Landau Institute for Theoretical Physics, 
   117940 Moscow, Russia

\end{center}

\vskip .3in

\begin{abstract}
The systematic approach for the calculations of the
non-perturbative contributions to the free energy
in the ferromagnetic phase of the random field Ising model
is developed. It is demonstrated that such contributions
appear due to localized in space instanton-like excitations
which exist only in dimensions $D \leq 3$. 
It is shown that away from the critical region such instanton 
solutions are described by the set of the mean-field saddle-point 
equations for the replica vector order parameter, 
and these equations can be formally reduced 
to the only saddle-point equation of the {\it pure system} 
in dimensions $(D-2)$. In the marginal case, $D=3$, the corresponding 
non-analytic contribution is computed explicitly. Nature of the phase 
transition in the three-dimensional random field Ising model 
is discussed. 
\end{abstract}

\vskip .1in


\vskip .3in

\section{Introduction}

Despite extensive theoretical and experimental efforts during the past
several decades very little is understood about 
the basic thermodynamic properties of the ferromagnetic Ising
systems with quenched random fields (for reviews see e.g. \cite{RFs}).
In the most simple form the random field spin Ising systems can be
described in terms of the Hamiltonian

\be
\label{rf1}
H \; = \; - \sum_{<i,j>}^{N} \sigma_{i} \sigma_{j} \; - \; \sum_{i} h_{i} \sigma_{i}
\ee
where the Ising spins $\{ \sigma_{i} = \pm 1 \}$ are placed in the vertices
of a D-dimensional lattice with the ferromagnetic interactions between
the nearest neighbors, and quenched random fields $\{ h_{i} \}$ are
described by the symmetric Gaussian distribution:

\be
\label{rf2}
P[h_{i}] \; = \;  \(2 \pi h_{0}^{2}\)^{-N/2} \;
\exp\[ - \frac{1}{2 h_{0}^{2}} \sum_{i=1}^{N} h_{i}^{2}\]  
\ee
where the parameter $h_{0}$ describes the strength of the random field

According to simple physical arguments by Imry and Ma \cite{Imry-Ma}
it should be expected that the dimensions $D_{c}$ above which the ferromagnetic
ground state is stable at low temperatures (it is called
the lower critical dimension) must be $D_{c} = 2$ (unlike the pure Ising
systems where $D_{c} = 1$)
Indeed, if we try to test the stability of the ferromagnetic state by flipping
the sign of the magnetisation in a large region of linear size 
$L$, we will find two competing effects: a possible gain of energy,
due go alignment with the random magnetic field, which scales as
$E_{h} = h_{0} L^{D/2}$; 
and the loss of energy, due to the creation of an interface,
which scales as $L^{D-1}$.
These estimates show that below dimensions $D_{c} = 2$ for any non-zero value
of the field $h_{0}$ at sufficiently large sizes $L$ the two energies are 
getting comparable, and therefore no spontaneous magnetization should be 
present. On the other hand, at dimensions greater than $D_{c} = 2$, the energy 
of the interface is always bigger than $E_{h}$. Therefore these 
excitations will not destroy the long range order, and a ferromagnetic 
transition should be present. These naive (but physically correct) arguments 
was later confirmed by a rigorous proof by Imbrie \cite{Imbrie}.

On the other hand, a perturbative study of the phase transition shows
that, as far as the leading large scale divergences are concerned, the
strange phenomenon of a dimensional reduction is present, such that
the critical exponents of the system in the dimension D appear to be the same 
as those of the ferromagnetic system without random fields in the dimension
d=D-2 \cite{D-2}. Since the lower critical dimension of the pure
Ising model is equal to 1, this result would imply that the lower critical 
dimension of the random field Ising model must be equal to 3, in
contradiction with the rigorous results. Actually, the procedure of summation
of the leading large scale divergences
could give the correct result only if the Hamiltonian in the presence of 
random fields has only one minimum. In this case the dimensional reduction
can be rigorously shown to be exact, by the use of supersymmetric arguments
\cite{Par-Sour-Par}.

However, one can easily see that as soon as the temperature is close enough
to the putative critical point (as well as in the whole low temperature 
region), there are local values of the
magnetic fields for which the free energy has more than one minimum
\cite{many-solu}.
In this situation there is no reason to believe that the
supersymmetric approach should give the correct results and therefore the
dimensional reduction is not grounded. Thus, the above arguments settle the
controversy about the lower critical dimension of the random field Ising
model in favour of the value $D_{c} = 2$.

Although at present we understand that at $D > 2$
the low temperature state of the RFIM must be
ferromagnetic, the nature of the phase
transition from the paramagnetic to the ferromagnetic state is still
a mystery. 
The only well established fact here is that the upper
critical dimensionality (the dimensionality
above which the critical phenomena are described by the mean-field
critical exponents) of such systems is equal to 6
(unlike pure systems, where it is equal to 4).
What is going on in the close vicinity of the phase transition, 
at dimensions $D < 6$ is not known.
And it is not just the question of what are the true values of the 
critical exponents. Even the nature of the phase transition
(is it of the second or of the first order) is nor clear. 
Moreover, at least in some cases there are indications
in favour of the existence of the intermediate spin-glass state
(with broken replica symmetry) separating paramagnetic and
ferromagnetic phases \cite{sg1,sg2}. 

The key problem here is the effect produced by the non-perturbative 
contributions which are coming from numerous local
minima states. In the present study, considering these contributions 
in the low temperature ferromagnetic phase (away from the critical point) 
at the {\it finite} strength of random fields $h_{0}$, 
I would like to "rehabilitate" the marginal character
of the dimensionality $D=3$ as well as of the trick of dimensional
reduction $D \to (D-2)$.
The point is that although numerous local minima configurations  
always exist in the ferromagnetic phase of the RFIM, they are not
always statistically relevant for the thermodynamic properties 
of the macroscopic system. In this paper I am going to demonstrate,
first in terms of  simple heuristic arguments (Section 2), and then using
more rigorous replica technique (Section 3), that
the non-perturbative states yield (non-analytic) contribution 
to the thermodynamics only in dimensions $D \leq 3$.
In terms of the qualitative heuristic arguments this 
contribution appears due to rare large cluster spin flips.
In terms of the non-perturbative replica formalism 
\cite{grif-dos,states-dos} such configurations are described by the instanton type
solutions of the mean-field saddle-point equations for the replica vector 
order parameter with broken replica symmetry.
These equations can be formally reduced to the only
saddle-point equation of the corresponding {\it pure system} in dimensions
$(D-2)$ (this fact has been noted first in \cite{Par-dos}), 
and then one can easily note that localized instanton-like
solutions formally exist only in dimensions
$D < 3$. In this sense the dimension $D = 3$ is marginal:
in dimensions $D > 3$ the non-perturbative states are irrelevant, while
in dimensions $D < 3$ they yield finite non-analytic contribution
(finally, when approaching the dimension $D=2$ from above
these states become so much relevant that they destroy the ferromagnetic
ground state of the system).
Similar instanton-like configurations 
in the presence of the homogeneous external magnetic field have been 
recently studied in Ref.\cite{mueller-silva}.
Although formally  the saddle-point equation in dimension $D = 3$ 
(in the zero external field) have no localized instanton solutions,
nevertheless, as it often happens in the marginal situations, the 
the relevant contributions can be taken into account in terms of
the instanton-like configurations containing the {\it soft-mode parameter} 
(which is the size of the instanton). 

Away from the critical point the concentration of the instantons 
(or the flipping clusters) is exponentially small, 
so that they can be treated as independent finite energy excitations. 
However, when approaching the putative phase transition point (from below) 
the typical distance between instantons eventually becomes
comparable with their size. In this situation
the present scheme of calculations breaks down, as 
the state of the system becomes just a mess of 
locally ordered "up" and "down" regions, when
it is impossible to separate the degrees of freedom
connected with the flipped clusters from those of the
ferromagnetic background. It is shown however, that at the {\it finite} strength
of the random fields this happens at temperatures where the system can still 
be described at the mean-field level, and it is argued that in this situation
it would be reasonable to expect the phase transition of the {\it first order}
into the disordered state (Section 4).

\section{Heuristic arguments}

\subsection{Perturbative contributions}

To compare the perturbative and the non-perturbative effects in the 
random field Ising model let us consider its 
Ginzburg-Landau continuous representation:
\be
\lb{rf3}
H\[\f({\bf x})\] \; = \; \I \Biggl[ \2 \(\n\f({\bf x})\)^{2}  
   + \2 \t \f^{2}({\bf x})  
   + \4 g \f^{4}({\bf x}) - h({\bf x}) \f({\bf x}) \Biggr] 
\ee  
Here $\t = (T/\T - 1)$ ($|\t| \ll 1$) is the reduced 
temperature parameter (for simplicity in what follows
it will be supposed that $\T =1$) and
the random fields $h({\bf x})$ are described by the 
Gaussian distribution,
\be
\lb{rf4}
P[h({\bf x})] = p_{0} 
\exp \Biggl( -\frac{1}{2h_{0}^{2}}\I \; h^{2}({\bf x}) \Biggr) \ ,
\ee
where $h_{0}$ is the parameter which describes 
the strength of the random field, 
and $p_{0}$ is an irrelevant normalization constant. 
The system will be considered in the low-temperature 
ferromagnetic phase, so that the reduced temperature 
parameter $\t$ will be taken to be negative,
$\t = -|\t|$. To neglect the effects
of the thermal fluctuations (away from the critical point),
it will be assumed that  the absolute value of the parameter $|\t|$
is not too small:  

\be
\lb{rf5}
|\t| \; \gg \; g^ {2/(4-D)} \; \equiv \t_{GL}
\ee
which is  the usual Ginzburg-Landau condition. 
The only relevant spatial scale in the system described by
the Hamiltonian (\ref{rf3}) is the correlation length:

\be
\lb{rf6}
R_{c}(\t) \; \sim \; |\t|^{-1/2}
\ee
(which under condition (\ref{rf5}) is described by the
mean-field critical exponent $\nu = 1/2$).
The statistical properties of the system at scales bigger
than the correlation length can be studied in terms of 
the saddle-point configurations  defined by
the equation

\be
\lb{rf7}
-\D\f({\bf x}) - |\t| \f({\bf x}) + g \f^{3}({\bf x}) \; = \; h({\bf x})
\ee  

In the absence of the random fields the ferromagnetic ground state of 
the system is described by the homogeneous configuration
$\f_{0} = \sqrt{|\t|/g}$, and it has the free energy density 
$f_{0} = - \t^{2}/(4 g)$.
In the usual perturbative approach the effects produced by the
random fields can be taken into account e.g. in the following way.
After rescaling the fields

\be
\lb{rf8}
\f({\bf x}) \; = \; \(\fr{|\t|}{g}\)^{1/2} \tl{\f}({\bf x}/R_{c}) ,
\ee
instead of eq.(\ref{rf3}) one obtains the system 
which is described by the rescaled Hamiltonian

\be
\lb{rf9}
H\[\tl{\f}({\bf z})\] \; = \; \fr{|\t|^{\fr{(4-D)}{2}}}{g} \int d^{D} z 
\Biggl[ \2 \(\n\tl{\f}({\bf z})\)^{2}  
   - \2 \tl{\f}^{2}({\bf z})  
   + \4 \tl{\f}^{4}({\bf z}) \Biggr]
   - \fr{|\t|^{\fr{(1-D)}{2}}}{\sqrt{g}}\int d^{D} z  \; 
   \tl{h}({\bf z}) \tl{\f}({\bf z})  
\ee 
where ${\bf z} \equiv {\bf x}/R_{c} = |\t|^{1/2} {\bf x}$. 
Here  rescaled random fields

\be
\lb{rf10}
\tl{h} ({\bf x}/R_{c}) \; = \; 
R_{c}^{-D} \int_{|{\bf x}'-{\bf x}|<R_{c}} d^{D} x' \; h({\bf x}')
\ee
are described by the distribution function

\be
\lb{rf11}
P[\tl{h}({\bf z})] = \tl{p}_{0} 
\exp\Biggl( -\frac{1}{2h_{0}^{2} |\t|^{D/2}}
         \int d^{D} z \tl{h}^{2}({\bf z}) \Biggr) \ ,
\ee
which is characterized by the mean square value

\be
\lb{rf12}
\ol{\tl{h}^{2}} \; = \; |\t|^{D/2} h_{0}^{2}
\ee
The ground state configurations of the fields $\tl{\f}({\bf z})$
are now defined by the saddle-point equation

\be
\lb{rf13}
-\D\tl{\f}({\bf z}) - \tl{\f}({\bf z}) + \tl{\f}^{3}({\bf z}) \; = \; 
   \fr{g^{1/2}}{|\t|^{3/2}} \tl{h}({\bf z})
\ee  
In the absence of the random fields ($\tl{h}\equiv 0$) 
the above equation would yield the trivial homogeneous 
ferromagnetic ground state solution $\tl{\f}({\bf z}) = 1$
(or $\tl{\f}({\bf z}) = -1$). In the presence of the random 
field term the ground state solution can be represented
as 

\be
\lb{rf14}
\tl{\f}_{gs}({\bf z}) = 1 + \psi({\bf z})
\ee
where $\psi({\bf z})$ describes small ($\psi \ll 1$)
spatial fluctuations of the ground state configuration.
Solving eq.(\ref{rf13}) in the {\it perturbative} way
(considering its r.h.s term as the small perturbation),
for  the typical value of these 
spatial fluctuations one easily finds:

\be
\lb{rf15}
\ol{\psi^{2}} \sim \fr{g h_{0}^{2}}{|\t|^{(6-D)/2}} + 
          O\( \(\fr{g h_{0}^{2}}{|\t|^{(6-D)/2}}\)^{2} \)
\ee
Substituting the solution eq.(\ref{rf14}) into the Hamiltonian
eq.(\ref{rf9}) for the free energy density of the ground state
configuration one gets:

\be
\lb{rf16}
f_{gs} = \fr{1}{V} H\[\tl{\f}_{gs}\] = 
- \fr{|\t|^{\fr{(4-D)}{2}}}{4g} \( 1 + \d \tl{f}\) \fr{1}{R_{c}^{D}} =
- \fr{|\t|^{2}}{4g} \( 1 + \d f\)
\ee
where $\d f$ is the random quantity with the typical value

\be
\lb{rf17}
\ol{(\d f)^{2}} \sim \fr{g h_{0}^{2}}{|\t|^{(6-D)/2}} + 
          O\( \(\fr{g h_{0}^{2}}{|\t|^{(6-D)/2}}\)^{2} \)
\ee
We see that the perturbation expansion goes in powers
of the (small) parameter

\be
\lb{rf18}
\fr{g h_{0}^{2}}{|\t|^{(6-D)/2}} \ll 1
\ee
which means that such perturbative approach is bounded
by the condition

\be
\lb{rf19}
|\t| \gg \t_{*}(h_{0}) = \(g h_{0}^{2}\)^{\fr{2}{6-D}}
\ee
On the other hand, according to eq.(\ref{rf9}), the typical value of the
{\it thermal} fluctuations 

\be
\lb{rf20}
\la\tl{\f}^{2}\ra \sim g |\t|^{-\fr{4-D}{2}}
\ee
In the pure system we can neglect the effects of the thermal fluctuations
provided $\la\tl{\f}^{2}\ra \ll 1$, which imposes the usual
Ginzburg-Landau condition, eq.(\ref{rf5}). Here,
in the presence of random fields, the thermal fluctuations can be 
neglected provided they are small not only compared to the average
value of the ferromagnetic order parameter (which in the present
notations is equal to one), but also compared to the typical value
of its random spatial variations 
(which is described by the field $\psi$, eq.(\ref{rf14})):

\be
\lb{rf21}
\la\tl{\f}^{2}\ra \ll \ol{\psi^{2}}
\ee
Substituting here eqs.(\ref{rf15}) and (\ref{rf20}) we find 
one more bound for the temperature parameter $|\t|$:

\be
\lb{rf22}
|\t| \ll h_{0}^{2}
\ee
Thus, taking into account eq.(\ref{rf19}), we conclude that 
the perturbative mean-field consideration of the 
present system is legitimate in the temperature interval

\be
\lb{rf23}
\(g h_{0}^{2}\)^{\fr{2}{6-D}} \ll |\t| \ll h_{0}^{2}
\ee
which automatically requires that the typical value of the
random field can not be too small:

\be
\lb{rf24}
h_{0}^{2} \gg g^{\fr{2}{4-D}}
\ee
Note that under this condition 

\be
\lb{rf25}
\t_{GL} \ll  \t_{*}(h_{0})
\ee
Another words, when approaching the phase transition from below
(which means reducing the absolute value of the temperature parameter
$|\t|$), first one arrives to the crossover temperature
$\t_{*}(h_{0})$ when the random fields can not be considered
as small perturbations any more (while the thermal fluctuations
can still be neglected), and only after that, at $|\t| \sim \t_{GL}$
(where the perturbation theory is no more valid)
the thermal fluctuations would become important.

In the particular case of the dimension $D = 3$ (which is the main focus
of our further study) the expansion parameter of the perturbation theory
is

\be
\lb{rf26}
\fr{g h_{0}^{2}}{|\t|^{3/2}} \ll 1
\ee
The crossover temperatures are

\bq
\lb{rf27}
\t_{GL} &=& g^{2}
\nn
\nn
\t(h_{0}) &=& \(g h_{0}^{2}\)^{2/3}
\eq
and the consideration is limited by the temperature interval

\be
\lb{rf28}
\t(h_{0}) \ll |\t| \ll h_{0}^{2}
\ee
which can exist provided the typical value of the random
fields is not too small:

\be
\lb{rf29}
h_{0} \gg g
\ee
Within these bounds the perturbative calculation of the free energy 
can be expressed in the form of eqs.(\ref{rf16})-(\ref{rf17}).

\subsection{Non-perturbative contributions}

Besides perturbative contributions
described above, one can observe completely different type of 
thermal excitations. Let us suppose that in the low
temperature phase the system has the global ferromagnetic 
magnetization directed "up", and let us consider 
a spatial island with sufficiently large linear size $L$,
in which the average value of the (random) field $\tl{h}$ is negative and  
sufficiently strong. Then one can easily note that in addition to the 
state "up" (with slightly modified value of the order 
parameter), another local minimum with orientation
"down" can be formed within this island.
Considering our system in terms of the rescaled Hamiltonian, 
eq.(\ref{rf9}), one can easily see that this alternative state 
can exist only if the gain in the energy due to the interaction with
the random field, 

\be
\lb{rf30}
E_{h} \sim - \fr{|\t|^{\fr{(1-D)}{2}}}{\sqrt{g}} L^{D} |\tl{h}| 
\ee
overruns the loss of energy due to the creation of the
interface, 

\be
\lb{rf31}
E_{f} \sim  +\fr{|\t|^{\fr{(4-D)}{2}}}{g}  L^{D-1} 
\ee
These estimates demonstrate that for a given value of $L$
such double-state situation in the considered island
is created provided 

\be
\lb{rf31a}
|\tl{h}|  >  h_{c}(L) \sim \fr{|\t|^{3/2}}{\sqrt{g} L}
\ee
which has a finite probability for any finite value of $L$.

Let us consider the statistical properties of such type of 
excitations in more detail. In terms of the Hamiltonian, 
eq.(\ref{rf9}), the energy of the spherical
region of the radius $L$ in which the magnetization is
flipped "down" can be represented as  
the sum of two contributions: 

\be
\label{rf32}
E(L) \; = \; (const) \fr{|\t|^{\fr{(4-D)}{2}}}{g} L^{D-1} - V(L)
\ee
The first term here 
is the positive energy due to creation of the spherical interface,
while the second contribution is the random energy which appear due to
the interaction with the random field, which in the continuous limit 
it can be represented as

\be
\label{rf33}
V(L) \; = \; 2 \fr{|\t|^{\fr{(1-D)}{2}}}{\sqrt{g}}  
          \int_{|{\bf z}|<L}d^{D}z \; \tl{h}({\bf z})
\ee
According to this definition,
\be
\lb{rf34}
\ol{V^{2}(L)} \; \sim \; \fr{|\t|^{(1-D)}}{g}  L^{D} \; \ol{\tl{h}^{2}} \; \sim \;
\fr{|\t|^{\fr{(2-D)}{2}}}{g}  L^{D} \; h_{0}^{2}
\ee
Since the typical value of the random energy $\[\ol{V^{2}(L)}\]^{1/2}$ scales as
$L^{D/2}$ we conclude that in dimensions $D > 2$ 
the first term in the total energy $E(L)$, eq.(\ref{rf32}),
must dominate at large scales, and therefore $E(L)$ 
is (on average) a growing function of $L$ (this is nothing else but 
the familiar Imry-Ma arguments \cite{Imry-Ma} which explain why the 
ferromagnetic state is stable at $D>2$). The point, however, is that
$E(L)$ is the {\it random} function, and it grows with $L$ only {\it on average}, 
while for a given realization of disorder it can have   
local minima at various (large) values of $L$. 

The probability distribution which describes the statistics
of the random functions $V(L)$ is given by

\be
\label{rf35}
{\cal P}[V(L)] \; = \; 
\int {\cal D}\tl{h}({\bf z}) \; \; P[\tl{h}({\bf z})] \;
\prod_{L} \left[\d\(\fr{2|\t|^{\fr{(1-D)}{2}}}{\sqrt{g}}  
          \int_{|{\bf z}|<L}d^{D}z \; \tl{h}({\bf z}) - V(L)\)\right]
\ee
where $P[\tl{h}({\bf z})]$ is the probability distribution function,
eq.(\ref{rf11}). Performing straightforward Gaussian integrations one obtains

\be
\label{rf36}
{\cal P}[V(L)] \; = \; (const) \;
\exp\left[ - \fr{g}{8 h_{0}^{2} |\t|^{\fr{(2-D)}{2}} S_{D}}\int_{1}^{\infty} \frac{dL}{L^{D-1}}
\left(\fr{dV(L)}{dL}\right)^{2} \right]
\ee
where $S_{D} = 2 \pi^{D/2}/ \G(D/2)$ is the square of the unite sphere
in $D$ dimension.
Since we are going to consider only large scales, 
the above result is given in the continuous limit
representation containing the ultraviolet
cut-off of the order of 
the correlation length, which in the present notations is equal to one.
We see, that in accordance with the physical meaning of the function $V(L)$,
its statistical distribution depends only on its derivative
(the constant term in this function is irrelevant), and therefore
the problem would become well defined only if we fix a value
of this function at a given $L$. For simplicity, let us assume
that at the lowest possible scale $V(L=1) =  0$. 
  
The above probability distribution function, eq.(\ref{rf36}), can be used 
to estimate the probability that the random function $E(L)$ has at least
one local minimum at scales larger than a given scale $L$.
Since the value of the energy $E(L)$ 
in a putative minimum growth with $L$, its probability 
must be small at large scales.
In this situation the sufficient condition for the existence
of a minimum somewhere beyond a given size $L$ is $\fr{dE(L)}{dL} < 0$,
or

\be
\label{rf37}
\frac{dV(L)}{dL} > (const) \fr{|\t|^{\fr{(4-D)}{2}}}{g} L^{D-2}
\ee
The probability $P_{min}(L)$ that the above condition is satisfied at a unit
length at the given size $L$ can be easily estimated using the general
distribution function (\ref{rf36}). Formally it can be obtained by 
integrating ${\cal P}[V(L)]$ over all functions $V(L)$ conditioned by eq.(\ref{rf37}). 
It is clear, however, that with the
exponential accuracy, the result of such integration is defined
by the lower bound for the derivative
$dV(L)/dL$. Therefore, substituting (at a given value of $L$)
$\frac{dV(L)}{dL} = (const) \fr{1}{g} |\t|^{\fr{(4-D)}{2}} L^{D-2}$
into eq.(\ref{rf36}), with the exponential accuracy one gets:

\be
\label{rf38}
P_{min}(L) \sim
\exp\[ -(const) \fr{|\t|^{\fr{(6-D)}{2}}}{g h_{0}^{2}} L^{D-3}\]
\ee

Thus we have derived very important property of the random  
function $E(L)$, eq.(\ref{rf32}):  
although, according to eq.(\ref{rf34}) 
this function (in dimensions $D > 2$) on average grows  
with $L$, the probability of finding a local minimum of this function in 
dimensions $D < 3$, according to eq.(\ref{rf38}), also grows with $L$.
The situation in the three-dimensional case
is marginal, and it is not quite clear to what extent 
the above simple arguments are grounded for $D=3$ (as usual in the marginal 
situations, more rigorous methods has to be used). Nevertheless, 
if we formally substitute $D=3$ into eq.(\ref{rf38}), 
we would have to conclude
that the probability of a local minimum becomes 
independent of the size of the flipped cluster.
Since the value of $P_{min}(L)|_{(D=3)}$ is exponentially small 
(in the parameter $|\t|^{3/2}/(g h_{0}^{2}) \gg 1$, eq.(\ref{rf26})),
the contribution to the free energy of such type of 
cluster excitations with the exponential accuracy is defined
by their probabilities (while their energies define a pre-exponential
factor).Thus we can estimate the non-perturbative part 
of the free energy density in the ferromagnetic phase of the 
3D random-field Ising model as

\be
\label{rf39}
\D f \; \sim \; \exp\( -(const) \fr{|\t|^{3/2}}{g h_{0}^{2}} \)
\ee
In the next section we shall re-derive this result in terms of the
systematic replica approach, which, in particular, allows to
calculate the $(const)$ factor.

\section{Non-perturbative replica approach}

The general scheme of the non-perturbative replica 
calculations has already been discussed in detail in the recent 
papers \cite{grif-dos,states-dos}.
Here we repeat it just in brief.
Let us consider a general  
$D$-dimensional random system 
described by a Hamiltonian $H\[\f({\bf x}); h({\bf x})\]$, 
where $\f({\bf x})$ is a field which defines
the microscopic state of the system, and 
$h({\bf x})$ are quenched random parameters.
Let us suppose that in addition to the
ground state, this system has another thermodynamically 
relevant local minima states located "far away"
from the ground state and separated from it 
by a finite barrier of the free energy.
In other words, it is  {\it supposed} that the partition 
function (of a given sample) can be represented in
the form of two separate contributions:

\be
\lb{rf40}
Z  = \int {\cal D}\f({\bf x}) \; \mbox{\large $e$}^{-\b H} \; = \; 
  \mbox{\large $e$}^{-\b F_{0}} + \mbox{\large $e$}^{-\b F_{1}} 
 \; \equiv \;Z_{0} \; + \; Z_{1}
\ee
where $F_{0}$ is the contribution coming 
from the vicinity of the ground state,
and $F_{1}$  is the contribution of the local minima states.
Then, for the averaged
over disorder total free energy we find:

\be
\lb{rf41}
{\cal F} = -\fr{1}{\b} \ol{\ln Z} 
= \ol{F_{0}} - \fr{1}{\b} 
\ol{\ln\[1 + Z_{1} Z_{0}^{-1}\]}
\ee
The second term in the above equation, 
which will be denoted by $\D F$, 
can be represented in the form of the series:

\be
\lb{rf42}
\D F = - \fr{1}{\b} \sum_{m=1}^{\infty}
\fr{(-1)^{m-1}}{m} \ol{Z_{1}^{m} Z_{0}^{-m}}
= - \fr{1}{\b} \lim_{n\to 0} \sum_{m=1}^{\infty}
\fr{(-1)^{m-1}}{m} Z_{n}(m)
\ee	      
where

\be
\lb{rf43}
Z_{n}(m) = \prod_{b=1}^{m} \int {\cal D}\f^{(1)}_{b}
         \prod_{c=1}^{n-m} \int {\cal D}\f^{(0)}_{c} \; 
\mbox{\Large $e$}^{ -\b H_{n}\[\f^{(1)}_{1},...,\f^{(1)}_{m}, 
           \f^{(0)}_{1},...,\f^{(0)}_{n-m}\]}    
\ee
is the replica partition function
($H_{n}\[{\boldsymbol\phi}\]$ is the corresponding
replica Hamiltonian), 
in which the replica symmetry in the $n$-component 
vector field $\f_{a}$ ($a=1,...,n$) is assumed to be 
broken. Namely, it is supposed that the saddle-point
equations

\be
\lb{rf44}
\fr{\d H_{n}\[{\boldsymbol\phi}\]}{\d \f_{a}({\bf x})} \; = \; 0
\;, \; \; \; \; (a = 1, ..., n)
\ee
have non-trivial solutions with the RSB structure

\be
\lb{rf45}
\f_{a}^{*}({\bf x}) = \left\{ \begin{array}{ll}
                 \f_{1}({\bf x})   & \mbox{for $a = 1, ..., m$}
		 \\
		 \\
                 \f_{0}({\bf x})   & \mbox{for $a = m+1, ..., n$}
                            \end{array}
                            \right.
\ee
with $\f_{1}({\bf x}) \not= \f_{0}({\bf x})$, so that the integration
in the above partition function, eq.(\ref{rf43}), goes over
fluctuations in the vicinity of these components.
It should be stressed that to be thermodynamically relevant,
the RSB saddle-point solution, eq.(\ref{rf45})
should be {\it localized} in space and 
characterized by  a {\it finite}
space size $R(m)$ and finite energies $E(m) = H_{n}\[{\boldsymbol\f}^{*}\]$. 
In this case the partition 
function, eq.(\ref{rf43}), will be proportional to
the entropy factor $V/R^{D}(m)$ 
(where $V$ is the volume of the system), and
the corresponding free energy contribution 
$\D F$, eq.(\ref{rf42}), will be extensive quantity.

Thus, to compute the non-perturbative free energy contribution
one should
find all saddle-point RSB solutions $\f_{a}^{*}({\bf x})$,
eq.(\ref{rf44}) and their corresponding 
energies $E(m)$ (in the limit $n\to 0$), and finally
one has to sum up the series

\be
\lb{rf46}
\D F = -\fr{V}{\b} \sum_{m=1}^{\infty} \fr{(-1)^{m-1}}{m}  
   R^{-D}(m) \; \(\b \det \hat T\)^{-1/2}_{n=0} \; 
\mbox{\Large $e$}^{-\b E(m)}
\ee
Here 

\be
\lb{rf47}
T_{aa'} \; = \; 
\fr{\d^{2} H\[{\boldsymbol\f}\]}{\d\f_{a} \d\f_{a'}}\Bigl|_{
{\boldsymbol\f}={\boldsymbol\f}^{*}}
\ee
and the Hessian pre-exponential factor $ (\det \hat T)^{-1/2} $
(taken in the limit $n \to 0$)
appears due to the integration over the replica fluctuations
in the vicinity of the RSB solutions, eq.(\ref{rf45}).
Note that in the present approach 
the procedure of analytic continuation  
$n \to 0$ is quite similar to that in the usual 
replica theory \cite{SG}: whenever the parameter
$n$ becomes an algebraic factor (and not the summation
parameter, or the matrix size, etc.), it can safely
be set to zero right away.

\vspace{5mm}

Coming back to the original Hamiltonian, eq.(\ref{rf3}),
and following the standard procedure,
after the Gaussian averaging of the replicated partition
function over the random fields $h({\bf x})$,
one obtains the replica Hamiltonian

\be
\lb{rf48}
H_{n}\[{\boldsymbol\f}\] \; = \;
 \I \Biggl[ \2 \sum_{a=1}^{n}\(\n\f_{a}\)^{2} 
      - \2 |\t| \sum_{a=1}^{n} \f_{a}^{2} 
      + \4 g \sum_{a=1}^{n} \f_{a}^{4} 
      - \2 h_{0}^{2} \sum_{a,b=1}^{n}  \f_{a} \f_{b}
   \Biggr]
\ee
The saddle-point configurations of the fields
$\f_{a}(x)$ are defined by the equations

\be
\lb{rf49}
-\D\f_{a}({\bf x}) - |\t| \f_{a}({\bf x}) + g \f_{a}^{3}({\bf x}) 
-h_{0}^{2} \(\sum_{b=1}^{n} \f_{b}({\bf x})\) = 0
\ee  
Below it will be demonstrated that besides the 
obvious (replica symmetric) ferromagnetic solution 
$\f_{a}({\bf x}) = \f_{0} = \sqrt{|\t|/g}$ these equations
have non-trivial localized in space
instanton-like solutions with the RSB 
structure:

\be
\lb{rf50}
\f_{a}^{*}({\bf x}) = \left\{ \begin{array}{ll}
\sqrt{\fr{|\t|}{g}} \; \psi_{1}({\bf x}\sqrt{|\t|}) &\mbox{for $a=1,...,m$}
		 \\
		 \\
\sqrt{\fr{|\t|}{g}} \; \psi_{0}({\bf x}\sqrt{|\t|}) &\mbox{for $a=m+1,...,n$}
                            \end{array}
                            \right.
\ee
These solutions are characterised by two non-trivial functions
$\psi_{1}({\bf z}) \not= \psi_{2}({\bf z})$ (where ${\bf z}={\bf x}\sqrt{|\t|}$).
Substituting these rescaled fields
 into the saddle-point eqs(\ref{rf49})
and into the Hamiltonian, eq.(\ref{rf48}), we find that
(in the limit $n\to 0$) the instanton configuration 
$\{\psi_{1}({\bf z}),\psi_{0}({\bf z})\}$ 
is defined by the equations

\bq
\lb{rf51}
-\D\psi_{1} - \psi_{1} + \psi_{1}^{3} - \lm(m) \(\psi_{1}-\psi_{0}\) &=& 0
\nn
-\D\psi_{0} - \psi_{0} + \psi_{0}^{3} - \lm(m) \(\psi_{1}-\psi_{0}\) &=& 0
\eq  
and its energy is

\be
\lb{rf52}
E(m) = m \fr{|\t|^{2-D/2}}{g}
\int d^{D} z \Biggl[ \2 \[(\n\psi_{1})^{2}-(\n\psi_{0})^{2}\] 
	        -\2 \[\psi_{1}^{2} -\psi_{0}^{2}\]
	        +\4 \[\psi_{1}^{4} - \psi_{0}^{4}\]
		-\2 \lm(m) \[\psi_{1} -\psi_{0}\]^{2}
		\Biggr] 
\ee
where

\be
\lb{rf53}
\lm(m) \; = \; \fr{h_{0}^{2} m}{|\t|}
\ee
We are looking for the localized in space
(spherically symmetric) solutions of the eqs.(\ref{rf51}),
such that the two functions $\psi_{1}(r)$ and $\psi_{0}(r)$
(where $r = |{\bf z}|$)
are different from each other in a finite region of space, 
and at large distances they both sufficiently quickly 
approach the same value $\psi_{0}=1$,
so that the integral in eq.(\ref{rf52}) will be converging.

Keeping in mind the qualitative arguments of the previous Section
it will be assumed that the parameters of the 
model satisfy the requirements, eqs.(\ref{rf23})-(\ref{rf25}). 
In this case the considered theory, eqs.(\ref{rf51})-(\ref{rf53}), 
contains the large parameter $\lm(m) \gg 1$ 
(for any $m = 1, 2, ...$), so that, according to eqs.(\ref{rf51}), 
the two fields $\psi_{1}$ and $\psi_{0}$ must be
close to each other. Redefining,

\bq
\lb{rf54}
\psi_{1}(r) \; &=& \; \psi(r) + \fr{1}{\lm} \chi(r)
\nn
\nn
\psi_{0}(r) \; &=& \; \psi(r) - \fr{1}{\lm} \chi(r)
\eq 
in the leading order in $\lm^{-1} \ll 1$ instead of
eqs.(\ref{rf51}) we get much more simple equations:

\bq
\lb{rf55}
-\D\psi - \psi + \psi^{3} - 2 \chi &=& 0
\nn
\nn
-\D\chi  + (3 \psi^{2} - 1) \chi &=& 0
\eq  
which {\it contain no parameters}. For the energy of
the configurations described  by the two fields
$\psi(r)$ and $\chi(r)$ instead of eq.(\ref{rf52})
(again, in the leading order in $\lm^{-1}$) 
we find the value, which does not depend on
the summation parameter $m$,

\be
\lb{rf56}
E \; = \; \fr{|\t|^{\fr{6-D}{2}}}{h_{0}^{2} g} \; E_{0}
\ee
where 

\be
\lb{rf57}
E_{0} \; = \; 
\int d^{D} z \[ (\n\psi)(\n\chi) 
             + (\psi^{3} - \psi)\chi - \chi^{2} \]
\ee
is the universal quantity which depends only on the 
dimensionality of the system.
Considering only spherically symmetric configurations,
eqs.(\ref{rf55}) can be represented as

\bq
\lb{rf58}
-\fr{d^{2}\psi}{d r^{2}} - \fr{D-1}{r} \fr{d\psi}{d r} + \psi^{3} - 2 \chi &=& 0
\nn
\nn
-\fr{d^{2}\chi}{d r^{2}} - \fr{D-1}{r} \fr{d\chi}{d r}  + (3 \psi^{2} - 1) \chi &=& 0
\eq  
Now as a matter of a simple algebraic exercise 
one can easily check that taking

\be
\lb{rf59}
\chi \; = \; - \fr{1}{r} \fr{d\psi}{d r}
\ee
the above two equations (\ref{rf58}) can be reduced
to one equation

\be
\lb{rf60}
-\fr{d^{2}\psi}{d r^{2}} - \fr{D-3}{r} \fr{d\psi}{d r} - \psi + \psi^{3}  = 0
\ee
for the only function $\psi(r)$. The corresponding energy, eq.(\ref{rf57}),
of the configurations described by eqs.(\ref{rf59}) and (\ref{rf60})
can be reduced to 

\be
\lb{rf61}
E_{0} \; = \; (D-2) S_{D}
\int dr \; r^{D-3}  \[ \2 \(\fr{d\psi}{d r}\)^{2} \; + \; \4 (\psi^{2} - 1)^{2} \]
\ee
Thus, we have reduced the problem of the non-perturbative
excitations in the $D$-dimensional random field Ising model to 
the study of similar instanton-like saddle-point configurations 
in the corresponding {\it pure} system in dimensions $(D-2)$.

Physically relevant solutions of the saddle-point equation
(\ref{rf60}) must be such that  in the limit $r \to \infty$ 
the function $\psi(r)$ sufficiently quickly approaches the value $\psi_{0}=1$
(so that the integral in eq.(\ref{rf61}) would be converging). 
It is evident, that at $D > 3$ (which corresponds to the "effective"
dimension $D_{eff} \equiv (D-2) > 1$) eq.(\ref{rf60}) has no such solutions. 
The formal solutions of this type and their contributions 
to the free energy in dimensions
$2 < D < 3$ (which correspond to $0 < D_{eff} < 1$) has been first 
considered in Ref.\cite{Par-dos},
and then studied in detail in Ref.\cite{states-dos}.

Here we are going to concentrate on the marginal case
$D = 3$, when eq.(\ref{rf60}) reduces to

\be
\lb{rf62}
\fr{d^{2}\psi}{d r^{2}} \; = \; \psi^{3} \; - \; \psi 
\ee
(where $0 \leq r < +\infty$).
Although this equation has no instanton
solutions, the configurations of the type

\be
\lb{rf63}
\psi(r) \; = \; \tanh\(\fr{r-L}{\sqrt{2}}\)
\ee
(which are the formal solutions of eq.(\ref{rf62}) for 
$-\infty < r < +\infty$) at large enough values of the
parameter $L$ can be considered as the soft-mode "quasi-instantons".
Using eq.(\ref{rf61})
one can easily find that at $L \gg 1$ the energy of such 
configuration is weakly
dependent on the soft-mode parameter $L$:

\be
\lb{rf64}
E_{0}(L) \; \simeq \; E_{*} \; - \; 8\sqrt{2} \pi \exp(-4 L/\sqrt{2})
\ee
where

\be
\lb{rf65}
E_{*} \; = \; \fr{8 \sqrt{2}}{3} \pi
\ee
is the energy of the infinite-size configuration.

It is clear that the energy $E_{0}(L)$ monotonously 
decreases with $L$, so that the 
true minimum is achieved when the above quasi-instanton
configuration completely disappears. 
Therefore the configurations
described by  eq.(\ref{rf63}) represents 
continuous spectrum of thermal excitations 
described by one parameter $L$. Correspondingly their 
contribution to the free energy is given by the 
summation (the integration) over all possible sizes $L$.
In terms of the original spatial notations
(${\bf x}={\bf z} |\t|^{-1/2}\equiv {\bf z} R_{c}$)
the size of the above quasi-instanton is
 
\be
\lb{rf66}
R \; = \; L \; R_{c}
\ee
Thus, coming back to the general expression for the
off-perturbative part of the free energy 
(with $D=3$ and $\b=1$)
eq.(\ref{rf46}), (where the energy $E(m)$  
and the size $R$ are defined in eqs.(\ref{rf56})
and (\ref{rf66})), and introducing here the integration over 
the soft-mode parameter $L$, we obtain

\be
\lb{rf67}
\D F = -\fr{1}{\b} \int_{1}^{\infty} dL \; \fr{V}{(L\R)^{3}} 
       \sum_{m=1}^{\infty} \fr{(-1)^{m-1}}{m}  \; 
       \exp\(-\fr{|\t|^{3/2}}{h_{0}^{2} g} E_{0}(L) \)
\ee
(here the ratio $V/(L\R)^{3}$ is the entropy factor which
yields the number of positions of the instanton with the 
linear size $(L\R)$ in the three-dimensional volume $V$).
Note that for the derivation of $\D F$ with the exponential accuracy,
in the considered range of parameters, eqs.(\ref{rf23})-(\ref{rf25}), 
the contribution of the fluctuations (contained in the term 
$(\det \hat T)^{-1/2}_{n=0}$, eq.(\ref{rf47}))
can be neglected \cite{states-dos}.   
Substituting here eq.(\ref{rf64}), and neglecting all
pre-exponential contributions, for the density of the 
non-perturbative part of the free energy 
we finally get (cf. eq(\ref{rf39}))

\be
\lb{rf68}
\D f \; \sim \;  
\exp\(- E_{*} \fr{|\t|^{3/2}}{g h_{0}^{2}} \)
\ee
where $E_{*} = \fr{8 \sqrt{2}}{3} \pi$.
Note that the validity of this result is limited by the condition:
$ (g h_{0}^{2})^{2/3} \ll |\t| \ll h_{0}^{2}$.

\section{Discussion}

The present study of the non-perturbative phenomena in the
random field Ising model has been done in terms of  
the mean-field local minima configurations in 
the low-temperature ferromagnetic state. The approach
completely neglects thermal fluctuations, which is justified, 
on one hand, provided the temperature is not too close to the 
putative phase transition point, and on the other hand, 
provided the temperature is not too low, 
where (weak) thermal fluctuations would nevertheless overrun
small spatial quenched fluctuations due to the interaction
with random fields. 
In terms of the Ginsburg-Landau Hamiltonian, eq.(\ref{rf3}),
in the dimension $D=3$
these two requirements impose the restrictions on the
value of the reduced temperature parameter,
$g^{2} \ll |\t| \ll h_{0}^{2}$, which automatically 
implies that the considered procedure of calculations
is formally valid only at {\it finite} strength of random
fields, $h_{0} \gg g$.  Under these restrictions
both the "hand-waving" heuristic arguments (Section 2.2),
and the formal replica calculations (Section 3) provide the
results, eqs.(\ref{rf39}) and (\ref{rf68}), which nicely fit each other.
Of course, the main thing here is not the actual value of the 
non-perturbative part of the free energy, but the physical mechanism,
which provides it. According to the speculations of
Section 2.2 the non-perturbative contributions appears due to
rare large cluster spin flips (which in terms
of replica calculations are described by the
localised in space instanton configurations). It is crucial
that such clusters (or the replica instantons) are supposed to
be far from each other, so that they can be treated as 
non-interacting and independent. According to the 
obtained results, eqs.(\ref{rf39}) and (\ref{rf68}), the spatial 
density of these clusters,
$\rho \sim \exp[ -(const) |\t|^{3/2}/(g h_{0}^{2})]$,
is exponentially small provided
$|\t| \gg (g h_{0}^{2})^{2/3} \equiv \t(h_{0})$.
In other words, when approaching the 
putative phase transition point from below (i.e. decreasing
the value of $|\t|$), at $|\t| \sim \t(h_{0})$,
the mean separation between clusters becomes 
comparable with their typical size. In this situation
the whole scheme of calculations breaks down, as 
the state of the system becomes just a mess of 
locally ordered "up" and "down" regions, when
it is impossible to separate the degrees of freedom
connected with the flipped clusters from those of the
ferromagnetic background. It is crucial that 
this happens at the temperature $\t(h_{0}) \gg \t_{GL} = g^{2}$
(well before the putative $\T$ of the expected
ferromagnetic/paramagnetic phase transition)
which is still far way from the Ginzburg-Landau crossover
temperature $\t_{GL}$ where the thermal fluctuations
would become important, and therefore here the system can still 
be described at the mean-field level.
Since at $|\t| \sim \t(h_{0})$
the local order parameter (the absolute value of the 
local magnetisation) is still finite, it would be 
reasonable to expect that somewhere at these temperatures 
the system undergoes the phase transition of the {\it first order}
into the disordered state. As for the nature of this
disordered state, it should be stressed that (unless the 
replica symmetry is broken!) one should not expect one more
phase transition from the spin-glass to the paramagnetic state.
The point is that due to the presence of quenched random fields 
the quantity $\ol{\la\f\ra^{2}}$ (which is the traditional 
replica-symmetric spin-glass order parameter)
remains non-zero at all temperatures. Therefore
one should expect not more than a crossover from "rather spin-glass"
(just after the transition) to "rather paramagnetic" (at high
temperatures) disordered states.

Unfortunately the analytic technique developed 
in this paper can not be directly applied for the 
description of this (rather exotic) disorder induced
first-order phase transition. Nevertheless, since the 
situation seems to stay at the mean-field level,
development of another analytic approach to this 
problem does not look completely hopeless.

\end{document}